\documentstyle[11pt,epsfig]{article}
\newlength{\capwidth}
\title{The D\O\ Detector at TeV33}
\author{The D\O\ Collaboration}

\begin{document}

\thispagestyle{empty}
\hfil{\parbox{6in}%

\begin{flushright}

{\large\bf D\O\ Note 3410}

\end{flushright}

\vspace{1.0in}

\begin{center}
{\Huge\bf The D\O\ Detector at TeV33 }
\end{center}

\vspace{0.5in}

\vspace{0.25in}

\begin{center}
\LARGE{\bf The D\O\ Collaboration}
\end{center}

\vspace{0.25in}

\begin{center}
\large{April 21, 1998}
\end{center}

\begin{abstract}

  The physics that can be done with 30~fb$^{-1}$ of data at the
  Tevatron (TeV33) will significantly advance our understanding of
  particle physics.  In this document we consider the potential of the
  D\O\ detector for running with the TeV33 Collider. In that era, the
  instantaneous luminosity is expected to rise by a factor of two to
  five.  We show that the D\O\ detector will perform well with
  some relatively modest modifications.

\end{abstract}

\vspace{ 0.5in}

\vspace{\fill}

\vspace{\fill}

\cleardoublepage

\tableofcontents
\cleardoublepage


\section*{Introduction}

 The D\O\ experiment commenced taking data at the Tevatron Collider
 during 1992 and continued through early 1996. D\O\ physics
 measurements~\cite{results} are prominent among the most important
 high energy physics results of this decade. Included among them are
 the observation of the top quark, the $W$ boson mass measurement,
 which is currently the best in the world, and pre-eminent measurements
 of couplings among the three electroweak bosons ($W$ and $Z$ and the
 photon). Many incisive measurements have been made of the (QCD) strong
 force. We have searched for new phenomena, such as SUSY
 particles, leptoquarks, compositeness effects and for deviations from
 the QCD description of strong interactions. While no breaches have
 been found in the armor of the standard model, these searches provide
 very important constraints on particle physics models.

 Building on the success of the original D\O\ design, in particular its high
 quality hermetic calorimetry, complete muon coverage, and the compact
 tracking volume, D\O\ is in the process of constructing an upgrade to the
 experiment~\cite{upgrade,comp}. This upgrade is to exploit the higher
 luminosity of the Tevatron in the era of the Main Injector. The major
 conceptual difference is in the tracking, 
 instrumented with a 2 Tesla solenoidal magnetic field, a fast
 scintillating-fiber tracker, and a comprehensive silicon tracking
 system. These components are characterized by short memory times, and are
 especially well suited to the high luminosity multibunch hadron collider
 environment. They also enhance the detector in a vital area of contemporary
 search physics, that of $b$-quark tagging. The upgraded detector will
 operate initially at luminosities of up to $2\times
 10^{32}$~cm$^{-2}$s$^{-1}$ and in the years 2000--2002 (Run~II) is expected
 to accumulate data corresponding to 2--4 fb$^{-1}$ of integrated
 luminosity. This corresponds to approximately 20-40 times the current data
 set.

 In what follows we discuss an exploration of the potential of the
 D\O\ detector as a platform for physics in a third era, that beyond
 Run II with the Main Injector. This era will be characterized by a
 mature collider complex, but one with much promise for further
 improvement. At that time, the Tevatron Collider will still be the
 highest-energy collider in the world, and the only machine capable of
 probing the most important aspects of high-$p_T$ physics: those
 touching on electroweak symmetry breaking and its manifestations.

 An upgraded Fermilab Tevatron (TeV33), capable of delivering
 integrated luminosities of order 30~fb$^{-1}$ to tape, offers a
 compelling physics program~\cite{DN}. The D\O\ experiment could
 expect to reconstruct approximately 30,000 $b$-tagged top
 decays~\cite{sm_frey}, to measure the $W$ mass to 20 MeV/{\it
 c}$^{2}$~\cite{sm_baur},
 and to pursue measurements in $b$ decays. The holy grail of such a
 program would be to discover supersymmetry and one or more Higgs
 bosons~\cite{sm_mrenna,sm_kim,sm_yao,sm_hedin}.

 In Section 1 we summarize the  evolution of the Tevatron
 Collider complex,  to define the possible operating
 parameters for the detector. Upgrades leading to utilization of an
 instantaneous luminosity of $2\times 10^{33}$~cm$^{-2}$s$^{-1}$ are
 possible~\cite{smass_tev33_1}. There are also scenarios (luminosity
 leveling) that could attain the desired integrated luminosity, but at
 reduced levels of instantaneous luminosity~\cite{sm_tev33_2}.

 In Section 2, we consider the performance  of the D\O\
 detector~\cite{upgrade} at luminosities up to $2\times
 10^{33}$~cm$^{-2}$s$^{-1}$ and integrated luminosities up to
 30~fb$^{-1}$. For several subdetectors, the relevant arguments have
 been presented in workshop or conference proceedings, and a summary
 of these is given in Ref.~\cite{sm_borch}. For other elements, the
 detailed aspects of the calculations appear in  appendices to this
 document.

 In Section 3, we consider modifications to the detector
 that would enhance its ability to handle the increased instantaneous
 and integrated luminosities.

 In Section 4, we draw from the considerations of the earlier sections
 to outline a detector that minimizes the changes required to exploit
 the physics potential of TeV33. Specifically, we assume that
 operation with luminosity leveling can achieve the integrated
 luminosity goals with the mean number of interactions not exceeding
 five per crossing.  While our conclusions are preliminary, it is clear
 that this or a similar upgrade to the D\O\ detector would lead to a
 superb physics program at the energy frontier in the years prior to
 initial exploitation of the CERN Large Hadron Collider (LHC).

\section{The TeV33 Collider}

 The possible upgrades to the Tevatron Collider complex are outlined
 in Ref.~\cite{smass_tev33_1}. The basic strategy is to increase the
 total number of antiprotons available for collisions. The present
 plans associated with the Main Injector project should lead to an
 instantaneous luminosity of $2\times 10^{32}$~cm$^{-2}$s$^{-1}$. The
 interbunch spacing will decrease from 396 ns to 132 ns at some point
 in Run II. With 396 ns bunch spacing (36 bunches of protons and of
 antiprotons), there will be about 5 interactions per crossing, while
 with 132~ns spacing the mean number is two, and these were in fact,
 the conditions assumed in the design of the D\O\ detector for Run
 II. There are several paths possible for improvements in D\O\ that
 would enable running at luminosities of the order of
 10$^{33}$~cm$^{-2}$s$^{-1}$.

 In addition to increasing the peak luminosity, it is possible to
 manipulate the machine parameters so as to start with a lower peak
 luminosity but to moderate the normal rate of fall in luminosity such
 that an integrated luminosity approximately 15\% below the maximum
 value can be attained~\cite{sm_tev33_2}. This technique, generically
 labeled {\it luminosity leveling}, can be achieved for example, by
 starting the store with an inflated $\beta^*$, and gradually reducing
 to the minimum $\beta^*$ ($\beta^*$ is the measure of focussing in
 the beams at the collision region, a high value means little
 focussing and a large beam, hence low particle density and low
 luminosity, conversely a low $\beta^*$ indicates high luminosity).
 This means that with a basic machine capability of $1\times
 10^{33}$~cm$^{-2}$s$^{-1}$, about 85\% of the maximum integrated
 luminosity can be obtained even for stores starting at $5\times
 10^{32}$~cm$^{-2}$s$^{-1}$.

 Another possibility that has similar intent is to increase the number
 of bunches in the machine. Following tradition, the machine with 132
 ns bunch spacing would still be operated with three abort gaps
 between the bunches, evenly spaced in the 21 $\mu$s circumference of
 the machine. This requires approximately 108 bunches of protons and
 of antiprotons. It is possible to operate with a single abort gap,
 which would increase the number of bunches to approximately 146, and
 for the same luminosity, provide a further reduction in the number of
 interactions per bunch.  The advantage of these gymnastics is
 self-evident: the detector experiences a mean number of no more than
 about five interactions per crossing.  In practice, it is likely that
 a combination of these measures would be employed. With machine
 performance at this level, operating scenarios can be constructed to
 reach an integrated luminosity $\sim$30~fb$^{-1}$ by ~2006--7.

\section{The D\O\ Run~II Detector: Limitations for TeV33 Operation}

  The D\O\ detector as implemented for Run~II of the Tevatron Collider
  has been described elsewhere~\cite{upgrade}. A side view of the D\O\
  experiment for Run~II is shown in Fig.~\ref{fig:d0layout}. In this
  section we concentrate on the limitations of the various subsystems
  for operation at the TeV33 Collider. These limitations fall into two
  categories: the deterioration of the detectors as a result of the
  increased integral radiation dose~(aging); and the limitation on
  detector and system operation as a result of the increased
  instantaneous luminosity and the concomitant increase in rates. In
  addition, we consider the effects of the environment on our ability
  to extract the crucial physics.

\begin{figure}[ht]
  \begin{center}
    \mbox{\epsfig{file=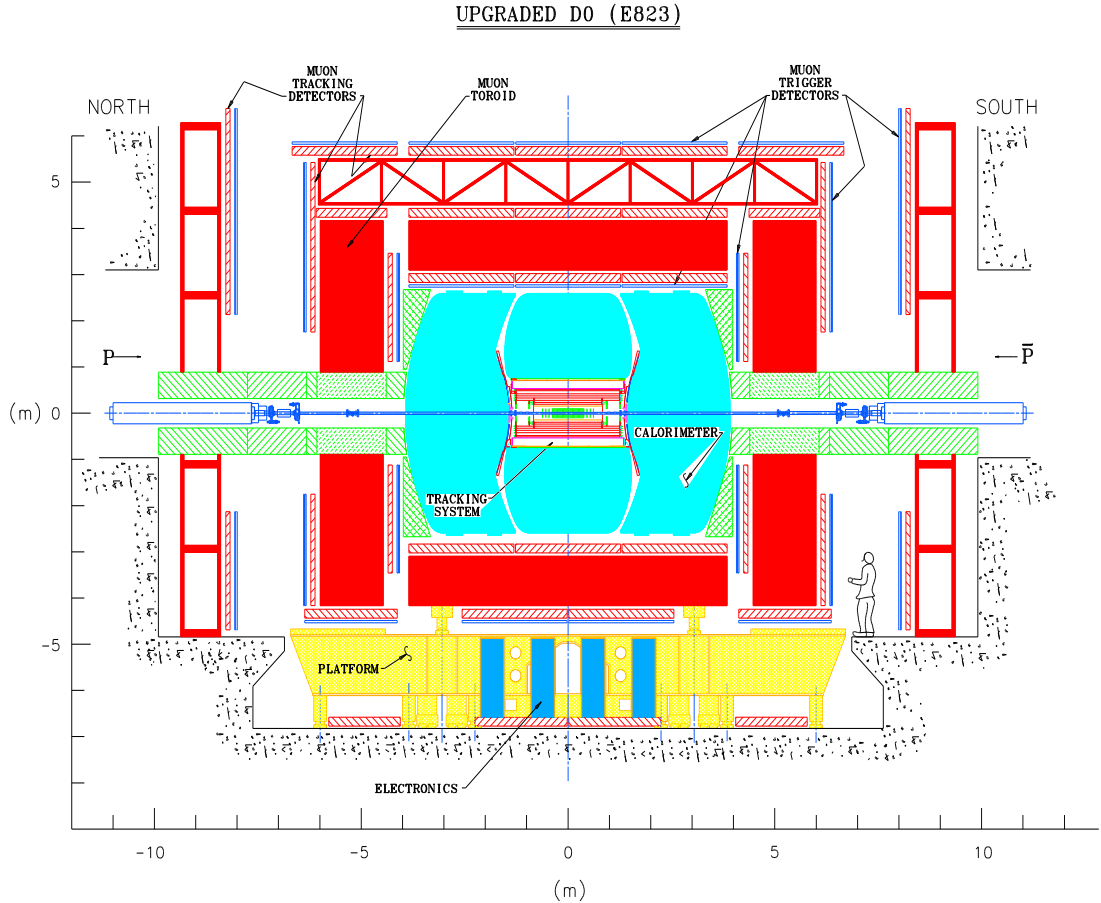,height=5 in,width=6 in}} \\
    \vspace*{0.5cm}
    \parbox{\capwidth}{%
    \caption{Layout of the D\O\ detector for Run~II.}
    \label{fig:d0layout}}
  \end{center}
\end{figure}

\subsection{Calorimeter}

  The D\O\ calorimeter consists of three cryostats, each containing
  calorimeter modules of depleted uranium, steel, and copper, immersed
  in liquid argon.  The electronics systems have been modified for
  Run~II to accommodate bunch crossing spacings of 132 ns. Although
  the technology is fundamentally radiation hard, the high
  instantaneous luminosity projected for TeV33 substantially increases
  the average energy deposited during the calorimeter's $\approx400$
  ns integration time. This increase leads to two principal areas of
  concern: the technical limitations on the calorimeter performance at
  high luminosity due to the calorimeter design, and the effect of the
  increased energy from underlying events on physics analyses.

\subsubsection{Technical Limitations}

As the luminosity is increased, the performance of the D\O\
calorimeter will be affected by the increase in energy absorbed per
unit time. While the D\O\ calorimeter is thought to be immune to
radiation damage, the increase in particle flux leads to an increase
in the ionization current flowing across the liquid-argon gap, thereby
dropping the gap voltage by a few percent at worst. 
This effect is discussed in Appendix A.

At present, the calorimeter readout  is expected to utilize three abort gaps in
the train of accelerator bunches. If the number of gaps were reduced
in order to increase the total number of bunches, the resultant small
increase in noise in the calorimeter would be acceptable (discussed below).

\subsubsection{Physics Effects}

The average energy deposited in a calorimeter tower ($0.1\times0.1$ in
$\eta-\phi$ space) by minimum bias events was measured during
Run~I using zero-bias triggers without noise suppression
(i.e., pedestal subtraction).  Scaling to a luminosity of $10^{33}$
cm$^{-2}$s$^{-1}$, with 108 bunches and the upgraded shaping
electronics, minimum bias events will contribute an average transverse
energy of $\approx 65$ MeV for central EM towers and $\approx 20$ MeV
for central hadronic towers.  For comparison, in Run~I the central
underlying-event energy was dominated by uranium noise, which gave an
average energy (not $E_T$) deposit of $\approx 5$~MeV in a central EM
tower and $\approx 9$ MeV in a central hadronic tower.  Thus, the
underlying event at TeV33 is expected to be dominated by the minimum
bias events accompanying the hard process.

The effect of underlying-event noise is expected to have minor impact
on most of the D\O\ TeV33 physics program.  The jet energy resolution
will continue to be dominated by fragmentation and showering
fluctuations.  The precision of the measurement of the top quark mass
is expected to be set by systematic errors in the jet energy scale and
the Monte Carlo modeling of gluon radiation.  A study of large missing
$E_T$ SUSY signatures~\cite{DN} found that the rejection of QCD
background events was not significantly affected by underlying-event
pileup.

The measurement that might be most affected by underlying-event noise
is the precision measurement of the $W$-boson mass.  The effect of increased
pile-up was studied for the TeV2000 report~\cite{2000W}, with the
conclusion that a $W$ mass uncertainty of 30 MeV/{\it c}$^2$ was
achievable with 10~fb$^{-1}$, using the currently favored
transverse-mass technique.  The resolution in transverse mass at high
luminosity is dominated by the resolution in missing $E_T$, with the
contribution from underlying-event noise becoming roughly equal to the
intrinsic detector resolution at a luminosity of $\approx 2\times
10^{32}$ cm$^{-2}$s$^{-1}$~\cite{1398}.  Further increases in
luminosity are likely to degrade the transverse-mass resolution.
Nevertheless, one would expect TeV33 to yield a factor of $\approx 10$
improvement in the statistical error over our present Run~I results.
The challenge is to achieve a comparable reduction in systematic
error.  Alternative techniques of measuring the $W$ mass are also
conceivable. For example, the transverse momentum distribution of the
lepton is less sensitive than is the transverse mass to degradation in
missing transverse energy resolution. One particular
approach~\cite{sm_srini} that has been tried is the ratio method, where
$W$ and $Z$ events are analyzed in an identical manner by ignoring one
of the electrons in $Z\to ee$ decay.  This method has been applied to
the D\O\ Run~Ia sample, yielding a statistical error of 360 MeV/{\it
c}$^2$ for 13~pb$^{-1}$.  The statistical error on the ratio method
extrapolates to $\approx 5-10$ MeV/{\it c}$^2$ for TeV33, with a
potentially smaller systematic error than can be achieved from fits to
the transverse mass. This is due to the reduced sensitivity to the
modeling of $W$ production, and the cancellation of effects from the
underlying event~\cite{Kotwal}. Note that this should be particularly
true for the muon decay channel.

\subsubsection{Summary}

The present D\O\ calorimeter appears to be capable of
operating in the TeV33 environment.  The increase in noise from
underlying events is mostly a concern for the precision measurement of
the $W$ mass. This measurement is expected to be limited by our
understanding of systematic uncertainties.

\subsection{Preshower Detectors}

 The D\O\ preshower detectors consist of a central unit~\cite{cps} and
 two forward modules~\cite{fps}. They are designed to improve electron
 identification in Run~II. The detectors~\cite{pscos} are composed of
 radiators and extruded triangular scintillating strips with embedded
 wavelength-shifting~(WLS) fiber readout. Scintillation light produced
 by charged particles in the radiators is collected by the WLS fibers
 and piped through clear fibers to visible light photon
 counters~(VLPC) outside the calorimeter. The central detector covers
 the pseudorapidity range $|\eta|<1.1$, and the forward detectors
 cover the range $1.4<|\eta|<2.5$. An online electron trigger is
 realized by requiring a large pulse in the preshower detectors to be
 matched with either a track in the fiber tracker for the central
 detector, or in the strips of scintillator in front of the converters
 in the forward detectors.

 In Appendix B, we discuss the potential problems from high occupancy,
 rate and radiation damage in operating the preshower detectors at
 TeV33. Occupancy will be approximately 10\% for an instantaneous
 luminosity of $5\times 10^{32}\ {\rm cm}^{-2} {\rm
 s}^{-1}$. Recent studies show that such rates do not have significant
 impact on the VLPC performance. The main effect is from
 radiation damage to the scintillator in high $|\eta|$ regions.
 At $|\eta|=2.5$, an integrated luminosity of 30 fb$^{-1}$ results in a
 dose of about 500 krad.

 We conclude that the central and part ($|\eta|<2.0$) of the forward
 preshower detectors can operate well for a TeV33 luminosity of
 $1\times 10^{33}\ {\rm cm}^{-2} {\rm s}^{-1}$.  The region of high
 pseudorapidity ($|\eta|>2.0$) is expected to suffer significant but
 gradual radiation damage. However, since most high-$p_T$ electrons
 are in the central region ($|\eta|<2.0$), the high-$p_T$ physics
 program is not expected to be affected by this problem.

\subsection{Muon Detectors}

 The layout of the D\O\ muon system for Run~II is shown in
 Fig.~\ref{fig:d0layout}. It consists of a central and two forward
 iron toroids and associated detectors.  In the central region
 ($|\eta|<~$1.0), two layers of scintillation trigger counters are
 used to trigger on muons and reduce cosmic ray
 backgrounds. Proportional drift tubes (PDT) are used to reconstruct
 muon tracks. They have drift cells with a cross section of
 5~cm~$\times$~10~cm and lengths of up to 6~m.

 In the forward region ($1.0<|\eta|<2.0$), there are three layers of
 scintillation counters with projective tower geometry. Their
 segmentation is 4.5$^\circ$ in $\phi$ by 0.1 in $\eta$.  For muon track
 reconstruction, there are three layers of mini drift tubes (MDT),
 with a total of ten detector planes.  They have drift cells of
 1~cm~$\times$~1~cm cross section, and are up to 6 m in length.  To
 reduce backgrounds from beam jets and interactions with accelerator
 elements, special shielding (see Fig.~\ref{fig:d0layout}) is
 installed around the beam pipe. It consists of soft steel,
 polyethylene, and lead, and it reduces background fluxes in the
 forward muon detectors by up to a factor 50 relative to the rates
 without shielding.  To accommodate the reduction in beam-crossing
 time from 3.5~$\mu$s to 132~ns, all of the muon electronics are being
 rebuilt using a deadtimeless pipeline.

 At a hadron collider, full geometric coverage for $|\eta|<$~2
 provides $\approx~$90\% acceptance for the decay products of massive
 objects.  The muon-detector acceptance for $|\eta|<$~2 is around
 80\%.  The losses are mainly due to the presence of ``no detector''
 zones under the calorimeter (supports) and between some chambers.

 A discussion of the operation of the muon system at higher
 luminosities can be found in Ref.\cite{smass_dima}.  Hits expected
 from muons are negligible in comparison to those from other sources.
 The most serious of these backgrounds is from remnants of high energy
 showers leaking through the detector.  In addition, low energy
 neutrons create a ``neutron gas'' inside the collision hall.  The
 only serious problem is the 15\% occupancy in the central PDT system
 for a luminosity of $1\times10^{33}~{\rm cm}^{-2}{\rm s}^{-1}$.

 Based on experience in running the detectors in 1994--96, the central
 muon PDTs can handle approximately 1~mC/cm of anode charge which
 corresponds to about 1 fb$^{-1}$ of integrated luminosity with the
 upgraded detector and improved shielding. The aging is primarily due
 to deposition on the signal wires. Although methods of cleaning anode
 wires have been developed, implementation requires a long shutdown
 period.  The useful lifetime between cleanings is much less than the
 30 fb$^{-1}$ exposure planned for TeV33, and this will consequently
 pose logistical difficulty.

In summary, all the muon detectors, except the central PDTs, will be able to
run at a luminosity of $1\times 10^{33}~{\rm cm}^{-2}{\rm s}^{-1}$,
and for an integrated luminosity of 30 fb$^{-1}$.

\subsection{Tracking Detectors}

 The D\O\ Tracker consists of a silicon-strip vertex tracker and a
 central scintillating-fiber tracker. The tracking devices are
 situated close to the beam (2.5~cm~$\leq r \leq$~52~cm) and surround
 the collision region, as shown in Fig.~\ref{fig:randy1}.  The silicon
 detector provides identification and reconstruction of primary event vertices
 and secondary decay vertices.  The fiber detector also provides a
 Level~1 track trigger with programmable $p_T$ thresholds.

\begin{figure}[ht]
\vspace{0.5in}
\centerline{\psfig{figure=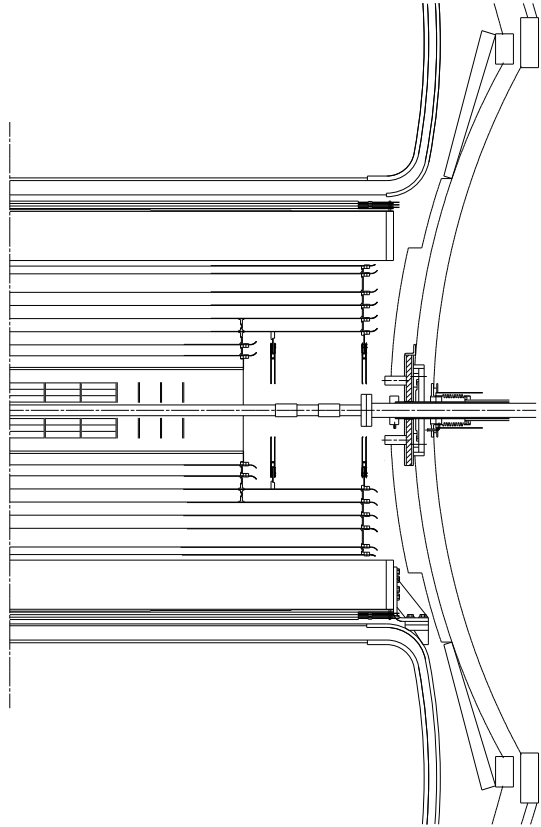,width=7.0in}}
\caption{ Elevation view of one end of the central region of the
upgraded D\O\ detector.  Shown at inner radii are three of the six
silicon barrels, each containing four layers of detectors, surrounded
by eight superlayers of scintillating fibers.  Additionally, there are
stations of silicon disk detectors covering the forward direction,
several of which are interspersed within the barrel region.}
\label{fig:randy1}
\end{figure}

 As the luminosity of the Tevatron collider rises above $2\times
 10^{32}$~cm$^{-2}$s$^{-1}$, the increase in the mean number of
 collisions per crossing ($\overline{n}$) leads to a rapid degradation
 of the Level 1 track triggering capability. Also, extended periods of
 high luminosity operation will pose a problem for detector
 performance for the inner layers of both detectors.

\subsubsection{Silicon Microstrip Tracker}

 Silicon devices suffer significant damage from large radiation
 doses~\cite{rad_damage}. Eventually this leads to a change in material
 properties. As a result of these changes, the voltage required for full
 depletion first decreases, and then increases as the material passes through
 type inversion, as shown in Fig.~\ref{fig:trig_5}.   As a result of the
 elevated depletion voltage, it is expected that the inner layers of the
 detectors will be inoperable after an integrated luminosity of the order of
 5--10 fb$^{-1}$.

\begin{figure}[ht]
  \begin{center}
    \epsfig{file=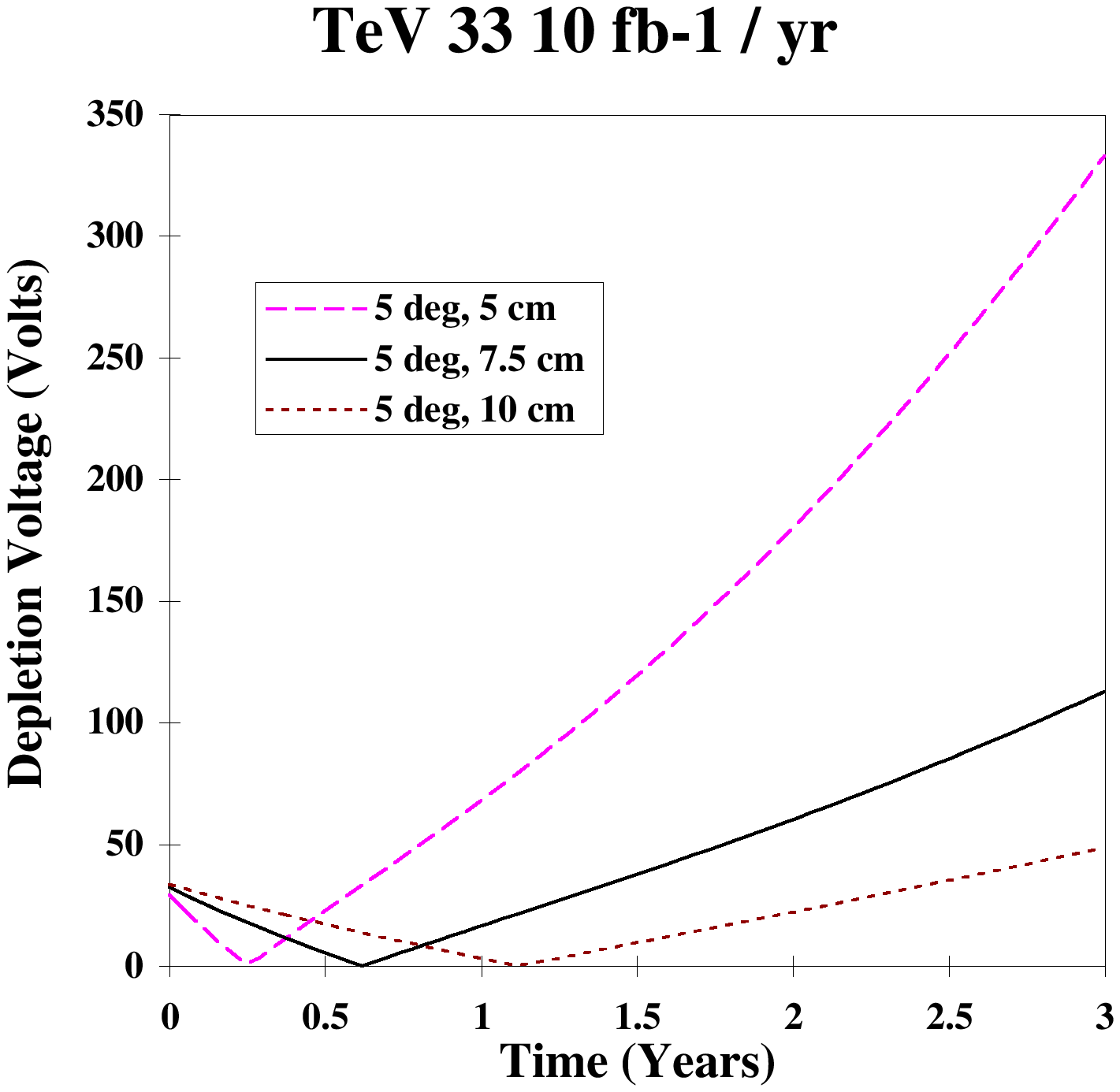,height= 9 cm}
    \caption{ Depletion Voltage vs. Time for layers 2, 3, and 4 of the
    Silicon Microstrip Tracker. }
    \label{fig:trig_5}
  \end{center}
\end{figure}

\subsubsection{Scintillating Fiber Tracker}

 The impact of high luminosity on the scintillating-fiber detector is
 discussed in Appendix C.  Higher luminosity operation at TeV33 will slightly
 modify the detected photoelectron
 distribution, fiber efficiency, and fiber-doublet ribbon
 efficiency. The primary problem is that occupancy will increase
 beyond 10\% for an instantaneous luminosity greater than $3-4\times
 10^{32}~{\rm cm}^{-2}{\rm s}^{-1}$. At this point, both the trigger
 performance and the pattern recognition capabilities will be put at
 risk.

\subsubsection{Summary}

 The inner layers of both the silicon detector and the
 fiber detector have problems with the increased luminosity of
 TeV33. In the case of the former, radiation damage will render the
 detectors inoperable. In the latter, as a result of the increased
 occupancy, the effectiveness of the detector will be much reduced.

\subsection{Triggers }

 On average, there are eighteen interactions per crossing at ${\cal L} =
 2\times 10^{33}~{\rm cm}^{-2}{\rm s}^{-1}$ and 108 bunches.
 Calorimeter-based triggers are not very sensitive to the number of
 minimum bias events in a crossing, but tracking triggers are
 sensitive because of the increased probability for false
 tracks. Trigger rejection for the fiber tracker decreases roughly
 quadratically with the number of minimum bias events so a factor of
 10 increase in luminosity worsens the rejection by a factor of
 100. In addition, the number of minimum bias events per crossing is
 Poisson distributed and, for example, when $\overline{n}$~=~18, one
 percent of the crossings will have 28 or more interactions.  Tracking
 triggers will unfortunately tend to select crossings with large
 numbers of minimum bias events rather than events of interest.

 We assume that TeV33 physics interests will center on processes with
 high $p_T$.  We expect a high $p_T$ trigger rate of 3--4 kHz in
 Run~II.  In a TeV33 scenario in which we employ luminosity leveling
 and use the maximum number of bunches, the event rate per crossing
 increases by a factor of up to four, so the real trigger rate
 increases approximately by this amount.  (A 4 kHz trigger rate is
 small compared to the 7 MHz crossing rate). This gives a Level 1
 trigger rate of 12--16 kHz. Although this does not greatly exceed our
 Run~II design goal of 10 kHz, the estimate contains no contingency. A
 factor of two would offer a reasonable safety factor, which would
 imply the need for a peak rate of about 20 kHz.

\subsubsection{Calorimeter Trigger}

 The present calorimeter Level 1 trigger forms towers that are
 0.2$\times$0.2 in $\eta$--$\phi$ space. Each tower is summed in depth
 to give an electromagnetic and a hadronic signal.  These signals are
 available to the trigger framework. In addition, signals from
 $4\times8$ ($\eta$--$\phi$) trigger towers are summed to form ``large
 tile'' signals which are used to search for jets.  There are four
 electromagnetic threshold reference sets, four jet threshold
 reference sets, and eight large-tile reference sets. Each set has
 different energy thresholds for each tower.  Several global
 quantities are summed for the entire calorimeter and also compared to
 sets of reference values. These quantities include the vector sum of
 transverse momenta (used for missing-$E_T$ triggers) and scalar sums
 of EM-$E_T$ and of Jet-$E_T$.  For Run~II, there is provision for
 direct readout to the DAQ system, and a scan of the history of the
 previous 25 crossings.

 For instantaneous luminosities in excess of $2\times 10^{32}~{\rm
 cm}^{-2}{\rm s}^{-1}$, the rejection offered by this system becomes
 inadequate.

\subsubsection{Muon Trigger}

 The estimated muon trigger rate at Level 1 for a luminosity of
 $2\times 10^{33}~{\rm cm}^{-2}{\rm s}^{-1}$, using the Run II trigger
 hardware is $>$~10 kHz (see Appendix D). The uncertainty on this
 number is large, reflecting uncertainties in our modeling accuracy
 and the limited Monte Carlo statistics presently available. However,
 with the expected TeV33 scenario of an instantaneous luminosity of
 $5\times 10^{32}~{\rm cm}^{-2}{\rm s}^{-1}$, this rate is
 approximately 4 kHz, or less with moderate tightening of muon
 requirements.

\subsubsection{Tracking Trigger}

The Level~1 tracking trigger for Run~II is formed by a logical AND of
hits from the 8 axial layers of the scintillating fiber tracker. Monte
Carlo studies show that the probability of finding a false track is a
steep function of the number of tracks in the detector. This is
demonstrated in Fig.~\ref{fig:randy7} where we display the increase in
wrong associations between different views in the fiber tracker. The
dependence on luminosity at different radii clearly shows that the
most critical effects are at the inner radii.

 \begin{figure}[ht]
\vspace{0in}
\centerline{\psfig{figure=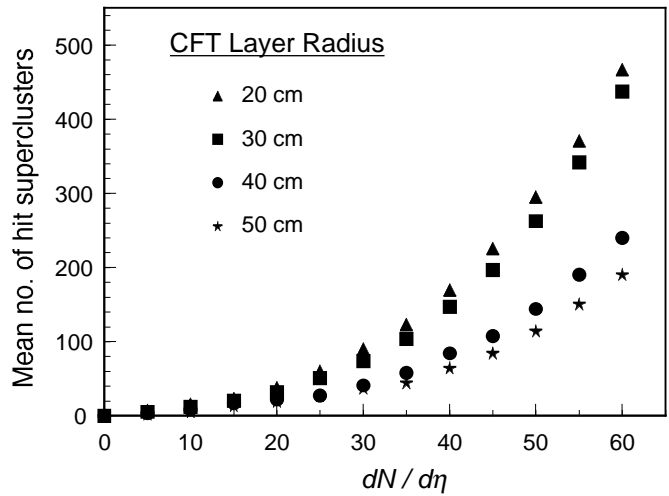,height=7.5in}}
\vspace*{-4.0in}
\caption{ Increase in fake track clusters as a function of track density, dN/d$\eta$. 
A dN/d$\eta$ of 50 corresponds to an $\overline{n}$~=~5, similar to 
luminosity-leveled TeV33 operation.  }
\label{fig:randy7}
\end{figure}

\section{The D\O\ Detector: Possible Upgrades}

\subsection{Calorimeter}

 We have seen that eliminating two of the three Tevatron abort gaps
 could allow an increase in the number of bunches from 108 to 146,
 with a consequent 26\% reduction in energy deposition in the
 calorimeter from underlying events.  Running with two less abort gaps
 will require some modification to the calorimeter readout since there
 are insufficient capacitors in the switched capacitor arrays (SCAs)
 to hold all 146 samples.  This modification may result in small
 pedestal shifts that will have to be corrected when the readout
 ``wraps'' around from the last capacitor in the SCA to the first.  In
 addition, there will be a slight increase in electronic noise due to
 longer times between baseline sampling.  The benefits  being
 independent of the particular bunch structure have led us to 
 incorporate the appropriate modifications for Run II.

\subsection{Preshower Detectors}

  A possible upgrade to the central preshower system is to replace the
  three layers of scintillating strips with one layer of scintillating
  pads, which would provide the tracker with the capability of triggering
  in both the $r$--$\phi$ and $r$--$z$ views.  A pad size of
  2~cm~$\times$~8~cm yields the same number of channels as the current
  design. The pad design effectively reduces the occupancy by a factor
  of four, and requires no additional photodetectors or electronics.

\subsection{Muon Detectors}

 High occupancies and fast aging require the replacement of the
 central proportional drift-tube system for TeV33.  We propose to
 replace these chambers with chambers similar to the MDTs.  The MDTs
 can be produced in any preselected length, in large quantities, and
 relatively inexpensively.  They can be made to be the same sizes as
 current PDT chambers, and will simply fit into the existing detector.
 A total of 72,000 channels is needed to replace the central
 system. (In the forward muon system, 48,000 MDT cells will be used
 for Run~II.) The rest of the muon system can be used unchanged.

\subsection{Tracking Detectors }

 An efficient central tracking system and a robust Level 1 trigger are
 important for successful operation of D\O\ under TeV33 conditions.
 Pattern recognition and tracking are great challenges at high
 luminosities. We base our considerations on the tracking system for
 Run II.  However, as discussed above, there are major problems for
 the inner layers of the silicon microstrip tracker being developed
 for Run II at the Tevatron, and for the inner layers of the
 scintillating-fiber tracker. Here we consider several recovery
 options.

 Since radiation exposure falls off nearly~\cite{cdfdosage} as the
 square of the radial distance from the beam, it is clear that
 withdrawing the inner silicon layers from 2.5 cm to a position at 4
 or 5 cm will lead to a factor of 2.5 to 4 increase in useful detector
 lifetime.  However, this can significantly degrade the
 impact-parameter resolution of the detector. 

 We can consider:

 \begin{itemize} 
 \item An R\&D effort to develop a suitable replacement detector:

    \begin{enumerate}

      \item Radiation hard silicon microstrip detectors. There are
      efforts under the auspices of the LHC R\&D program to develop
      such detectors with improved immunity to radiation~\cite{rose}.

      \item Development of appropriate silicon pixel
      detectors. Because of their smaller element size, these devices
      are intrinsically more radiation hard~\cite{heijne}.

      \item Development of an alternative strip technology, such as
      diamond~\cite{diamond}.

    \end{enumerate}

   \item Relocation of the silicon microstrip tracker to a larger
   radial region (4~cm~$\leq r \leq$~16~cm).

   \item Introduction of ``wide strip" silicon tracking and track
   triggering structures in the intermediate radial region 20~cm~$\leq
   r \leq$~30~cm~\cite{wattsjohnson,cdf-isl}.

   \item Modification of the fiber tracker:

   \begin{enumerate}

       \item Relocation of the scintillating-fiber tracking layers to
       larger radii (35 cm $\leq r \leq$ 52 cm). This has been
       studied, especially in connection with the
       trigger~\cite{wattspartridge}.

       \item Splitting the fibers in the middle (at $\eta$ = 0) for
       the layers at small radii. This increases the number of
       channels, and therefore also the space required to accommodate the
       clear connecting fibers and VLPC cryostats.

       \item Reduction of the fiber diameter from 830 $\mu$m to 500
       $\mu$m. R\&D on smaller diameter VLPCs would be required, and
       it may be that the increased number of channels would lead to
       space problems in extracting the signals. Some work might also
       be required on optimizing 500 $\mu$m fibers, although the
       projection from the performance of 830 $\mu$m is promising.

   \end{enumerate}

\end{itemize}

   Detailed simulation of tracking and triggering and an active
   ongoing program of detector R\&D are required to make a rational
   choice among these options, which are not necessarily exclusive.
   For the scintillating fibers this program includes studies and
   optimization of smaller-diameter scintillating fibers (500 $\mu$m),
   and smaller diameter VLPC pixels ($4\times8$ arrays in the same
   silicon area as the current $2\times4$ arrays).  For microstrip
   detectors, evaluation of diamond microstrip detectors, development
   of radiation-hard silicon strip detectors, and development of local
   trigger towers based on these technologies are needed.  For pixels,
   R\&D on adapting the LHC~\cite{heijne} pixel detectors and readout
   for D\O\ is required. (A Fermilab development effort~\cite{pixrd}
   has already been established in this last area).

\subsection{Triggers}

 It is necessary to address the increased Level~1 acceptance rate. We
 could move some of the rejection power in the Level~2 processors to
 Level~1, so that we can fit all the high $p_T$ physics into the
 existing 10 kHz bandwidth.

\subsubsection{Calorimeter Trigger}

 Improving calorimeter-trigger rejection requires a new Level 1
 trigger system for the calorimeter. This system would move the
 isolation criteria for EM jets to Level 1. It would improve the
 hadron part of Level 1 by allowing a more flexible combination of
 towers than is provided by the large tiles. That is, one could have a
 moving window that would search for isolated jets rather than large
 fixed tiles. This would also provide the $\phi$ of the trigger so
 that there could be a $\phi$ match between the tracker, preshower and
 calorimeter.  Putting the isolation cut at Level~1 should increase
 rejection by a factor of two to three for electrons.  It is not yet
 known if the changes to the hadron trigger would give the additional
 rejection that is required.

\subsubsection{Tracking Triggers}

 Three options have been considered. 

\begin{itemize} 

 \item An option~\cite{wattsjohnson} which involves a
 complete rework of the tracking to incorporate an extensive pixel
 system in a new technology, possibly microgap chambers.

 \item An option which goes hand in hand with a reconfiguration of
 fiber tracking layers to larger radii~\cite{wattspartridge} could
 provide more axial layers in the trigger than the Run II design.

 \item A novel system~\cite{maninarain} based on using a
 silicon pixel detector in the trigger at Level 1.

\end{itemize}

The first option is revolutionary and requires very extensive
modification of the tracking.  The second option is less powerful and
indeed would likely not be adequate for a luminosity of $ 2\times
10^{33}~{\rm cm}^{-2}{\rm s}^{-1}$. However it might be adequate in
the TeV33 luminosity-leveled scenario with an effective ${\cal L} =
5\times 10^{32}{\rm cm}^{-2}~{\rm s}^{-1}$. The third option depends on the pixel development.

\section{Conclusions}

 Based on the discussion of the previous section we advance a
 preliminary suite of upgrades, which would be sufficient to
 ensure good operation of the detector at luminosities up to and
 including a maximum instantaneous intensity of ${\cal L} = 5\times
 10^{32}~{\rm cm}^{-2}{\rm s}^{-1}$ with a bunch separation of 132
 ns.

\begin{itemize}
 
 \item{\bf Calorimeter} 

 No change.

 \item{\bf Preshower Detectors} 

 No change.

 \item{\bf Muon Detectors}

 Replace the central PDT system with the chambers similar to those that will be
 used in the forward muon system in Run~II.

 \item{\bf Tracking Detectors}

  \begin{enumerate}

   \item At small radii (less than 10 cm) introduce pixel detectors
   as replacements for the inner layers of the silicon detector.

   \item Between 10 cm and 30 cm in radius, augment the present silicon
   layers with new detectors, perhaps utilizing the space of the
   present inner scintillating-fiber layers.

    \item For radii greater than 30 cm, reconfigure the existing fiber
    tracker channels to maximize effectiveness in momentum
    determination and triggering.

  \end{enumerate}

 \item{\bf Triggers}

  \begin{enumerate}

     \item Base the Level 1 tracking trigger on the redeployment of
     the scintillating fiber tracker at larger radii, perhaps with a
     pixel trigger.

     \item Upgrade the Level 1 calorimeter system to use dynamic
     large-tile triggering, along with isolation for electromagnetic
     triggers.

  \end{enumerate}

\end{itemize}

 The physics that can be done with 30~fb$^{-1}$ of data at the
 Tevatron is highly compelling.  The D\O\ detector will perform well in the
 TeV33 environment with these or similar upgrades.


\clearpage

\section*{Appendices}

\appendix

\section{Rate Effects on the Calorimeter High Voltage  }

 We consider the effects of increased energy deposition rate on the
 D\O\ calorimeter operation. In particular we consider the effect of
 increased current draw on the effective high voltage and the
 consequent effect on the observed signal. We find the effect to be
 finite and manageable.

 The high voltage supply is connected to the resistive coat on the
 signal electrode planes. The ionization current was calculated as a
 function of pseudorapidity for the four electromagnetic(EM) layers by
 converting the average energy deposit to the average charge measured
 by the calorimeter electronics and taking into account the unmeasured
 image charge associated with the slow-moving positive ions. The
 ionization current was assumed to be evenly distributed among the
 liquid argon gaps in ganged subsection and the voltage drop across
 the resistive coat was calculated.  Since the drift velocity of the
 ionized electrons in the argon depends on the electric field, a large
 voltage drop across the resistive coat will change the shape of the
 signal pulse and decrease the charge collection efficiency.  The
 largest voltage drop occurs for the EM2 layer in the end
 calorimeter(EC) at largest $|\eta|$, located near the shower maximum
 for low $p_T$ photons from minimum bias events.

 The voltage drop has been calculated from the average energy per beam
 crossing deposited in the calorimeter.  The data used were taken  from zero
 bias runs  with calorimeter noise suppression turned off.  The
 ionization current is given by \begin{equation} I_{0} = \frac{\bar
 EF_S e f}{2I} \end{equation} where $\bar E$ is the average energy deposited in
 a beam crossing, $F_S$ is the sampling fraction, $e$ is the electron
 charge, $f$ is the beam crossing frequency, and $I$ is the ionization
 potential for argon.  
 
 The changing drift velocity will affect the calorimeter energy
 response and likely require luminosity- and pseudorapidity-dependent
 corrections. If this voltage drop is converted to a change in
 response, the effect is quite small, due to the rather gentle
 dependence of the response on high voltage in the plateau
 region. This is shown in Fig.~\ref{fig:calsag} where we have assumed
 an instantaneous luminosity of $5\times 10^{32}~{\rm cm}^{-2}~{\rm
 s}^{-1}$ and the current 2.0 kV operating value.

\begin{figure}[ht]
\centerline{\psfig{figure=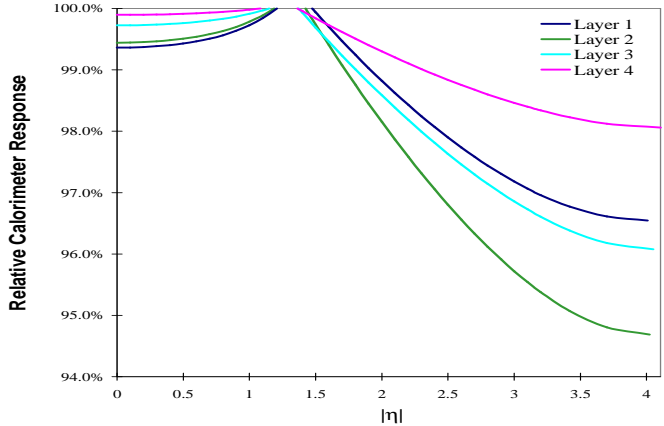,width=4.25in,height=3.75in}}
\vspace{-0.3in}
\caption{EM calorimeter response factor compared to zero luminosity  versus $|\eta|$ taking account the gain reduction and the change in drift speed as the effective high voltage is reduced due to current draw for 2.0 kV and ${\cal{L}} = 5\times10^{32}~{\rm cm}^{-2}~{\rm s}^{-1}$. The region 1.1 $< |\eta| <$ 1.4
has no EM calorimeter coverage.}
\label{fig:calsag}
\end{figure}


\section{Radiation Damage, Occupancy, and Rate Effects in the Preshower Detectors}

\subsection{Radiation Damage}

Albedo neutrons from the calorimeter, and charged and neutral hadrons
from $p\bar{p}$ interactions are the main sources of radiation in the
preshower detectors~\cite{dose}.

The central tracking volume is filled with albedo neutrons which are
uniformly distributed in $r$, the distance perpendicular to
the beams.  The logarithmic energy spectrum peaks roughly around
1~MeV. The estimated neutron flux density $\phi_n=1.2\times 10^4\
{\rm cm}^{-2} {\rm s}^{-1}$ at a luminosity of ${\cal L}=10^{32}\ {\rm
cm}^{-2} {\rm s}^{-1}$.  Using a conversion factor of
1~neutron/cm$^2=1.8\times 10^{-9}$~rad, the dose rate from neutrons is
given by $\frac{dD_n}{dt}=2.2\times 10^{-5} 
({\cal L}/10^{32}{\rm cm}^{-2}{\rm s}^{-1})$~rad/s.

Assuming a minimum bias  cross section of 50~mb, and 4 charged
particles per unit of pseudorapidity, the charged particle flux at a
radius $r$ is $\phi_\pm = 3.2\times 10^6/(r/1~{\rm cm})^2\ {\rm
cm}^{-2}{\rm s}^{-1}$.  Taking into account a factor of two for low
$p_T$ loopers, the dose rate from the charged particles is given by
$\frac{dD_\pm}{dt}=\phi_\pm \frac{1}{\rho}\frac{dE}{dx} =
\frac{0.2}{(r/1~{\rm cm})^2}$~rad/s, where $\frac{1}{\rho}\frac{dE}{dx}$ is the specific ionization.

Most neutral hadrons from beam interactions are $\pi^0$'s, with an
average $p_T$ of 0.5~GeV/c.  From isospin symmetry, the $\pi^0$ flux
is expected to be half the flux of charged particles. Assuming each
photon from $\pi^0$ decay cascades to 4 electrons in the solenoid and
the lead, the expected dose rate from neutral hadrons is
$\frac{dD_0}{dt} = 4\phi_\pm (\frac{1}{\rho}\frac{dE}{dx})
=\frac{0.6}{(r/1~{\rm cm})^2}$~rad/s.

Adding the rates from neutron, charged and neutral hadrons together,
the total dose rate at ${\cal L}=10^{32}\ {\rm cm}^{-2}{\rm s}^{-1}$
is $\frac{dD}{dt}=2.2\times 10^{-5} + \frac{0.8}{(r/1~{\rm
cm})^2}$~rad/s.  The total radiation doses for the central and forward
preshower detectors are shown in Fig.~\ref{fig:cps_rad} as functions
of integrated luminosity.  The detailed calculation can be found in
Ref.~\cite{dose}.  The above calculation has been checked using fully
simulated minimum bias events for the central detector. The numbers
from the calculation and the simulation agree very well.  The expected
radiation level for an integrated luminosity of 30~fb$^{-1}$ in
Run~III is 54~krad for the central preshower, 140~krad at $|\eta|=2.0$
and about 400~krad at $|\eta|=2.5$ for the forward preshower. The dose
in the central detector is well below the level at which the
scintillator or fibers are affected. However, the dose in the forward
detectors is high.  The scintillator and fibers in the region
$|\eta|>2.0$ will suffer gradual but substantial radiation damage and
therefore the performance of the forward preshower detectors will
degrade significantly over the last part of TeV33.

\begin{figure}[htbp]
  \begin{center} \mbox{\epsfig{file=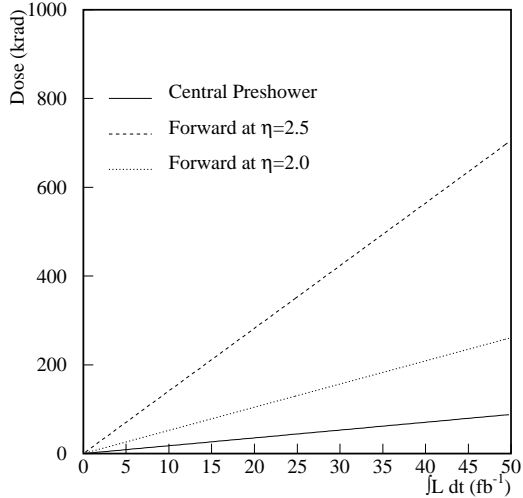,width= 7 cm}} \\
    \vspace*{0.5cm} 
    \parbox{\capwidth}{ 
    \caption{The total
    radiation dose in the central and
    forward preshower detectors at $|\eta|=2.0$ and 2.5, versus integrated luminosity.}  
    \label{fig:cps_rad}
    }
    \end{center}
\end{figure}

 \begin{figure}[h]
\begin{center}
\mbox{\epsfig{file=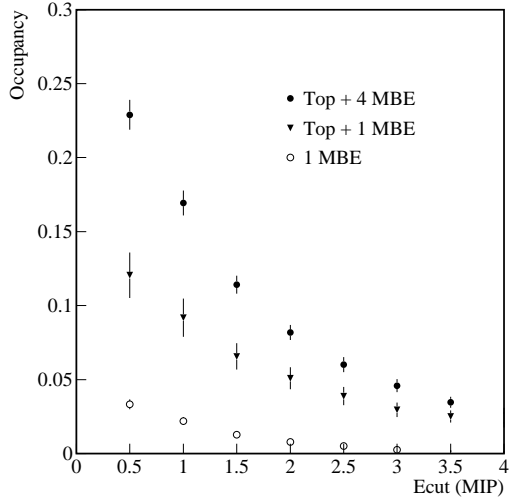,width=7 cm}}
\vspace*{0.5cm}
\parbox{\capwidth}{%
\caption{The expected occupancy as a function of the energy
deposition in the axial strips of the central preshower detector for minimum 
bias events (MBE) and for $t\bar{t}$ events with one and four minimum bias 
events overlayed.}
\label{fig:psbusy}}
\end{center}
\end{figure}

\subsection{Occupancy}

At a luminosity of ${\cal L}=1\times 10^{33}\ {\rm cm}^{-2} {\rm
s}^{-1}$ with a 132~ns bunch spacing, on average there will be about
7-8 interactions per crossing. Figure~\ref{fig:psbusy} shows the
expected occupancies as functions of the energy deposition for the
axial strips of the Run~II central preshower detector for minimum bias
events and $t\bar{t}$ events with four minimum bias events overlayed.
Here the occupancy is defined as the ratio of the number of strips
with hit with energies above threshold, and the total number of
strips. The occupancies for the stereo strips are expected to be
slightly worse. The occupancies at 1~MIP threshold are about 2\% for
one minimum bias event and about 17\% for a typical $t\bar{t}$ event
overlayed with four minimum bias events, where the unit MIP is defined
as the most probable energy deposited in the layer by a muon.
Assuming the occupancy increases linearly with the number of minimum
bias events, about 23\% of the strips will be hit for a $t\bar{t}$
event at a luminosity ${\cal L} =1\times
10^{33}$~cm$^{-2}$~s$^{-1}$. Even with a 3~MIP threshold, the
occupancy will be around 9\%. This high occupancy level will probably
affect the online electron trigger more than offline electron
identification.

\subsection{The Rate (Non)Issue}

The high rate expected for the high luminosity is a concern for the
visible light photon counters(VLPC). The issue of the gain and quantum
efficiency~(QE) of the VLPC at high rate has been studied in
detail. The single photoelectron~(PE) rate is approximately given by
$R = 2\cal{L}\sigma{\cal{O}}${\it y}, where ${\cal L}$ is the
instantaneous luminosity, $\sigma$ is the total cross section,
$\cal{O}$ is the occupancy with a 1~MIP threshold and ${y}$ is the
photoelectron yield for a 1~MIP energy deposition. The factor of two
is included to take into account the yield from hits below threshold.
Although failing the threshold, these hits still contribute to the
single PE rate. For the Run~II design luminosity of $ 2\times
10^{32}$~cm$^{-2}$s$^{-1}$, the single PE rate is about 8~MHz assuming
$\sigma=50$~mb, $\cal{O}$~=~2\% and ${y}=20$~PE/MIP~\footnote{This is
the expected photoelectron yield based on the cosmic--ray test,
see~\cite{pscos} for details.} for axial strips of the central
detector.  At a Run~III luminosity of $1\times
10^{33}$~cm$^{-2}$s$^{-1}$, the single PE rate is 40~MHz, five times
higher. However, it should be noted that a yield of 20~PE/MIP has a
considerable safety margin for achieving the required performance of
the preshower detector. Halving the yield
is not expected to significantly degrade the performance of the
detector. Nominally, a 20~MHz rate capability of the VLPCs is needed.
Including a factor of two for the uncertainty, we conclude that a
40~MHz VLPC rate capability will be sufficient for the Run~III
application. Recent studies (see Appendix C) of the VLPC
characteristics seem to indicate that the gain and QE of the new
VLPCs show little dependence on the rate implying that rate is not a
problem.


\section{ Radiation Damage, Rate Effects, and Occupancy in the Fiber Tracker}

Figure~\ref{fig:randy2} shows the mean detected photoelectron(PE)
signal for a scintillating fiber on the inner barrel of the CFT. The
multiclad fiber (830~$\mu$m diameter and 1.7~m length) is optically
coupled to a clear fiber waveguide of identical diameter and 10~m
length, and is read out with a VLPC. The minimum detected PE signal
occurs for minimum ionizing particles traversing the fiber at normal
incidence near the mid-point along the fiber length, which corresponds
to the minimum path length.  At this location ($\eta\sim 0$), we
expect to detect a mean of 11 PE/MIP. A threshold in photoelectron
equivalents is set for every channel of the CFT, which establishes the
noise rate and efficiency for each.  Typically, thresholds are set at
1--1.5 PE, yielding noise rates $<$ 0.5\%, and efficiencies$>$99\%.
The efficiency of a fiber-doublet ribbon also exceeds 99\% under these
conditions.
 	
\begin{figure}[htp]
\centerline{\psfig{figure=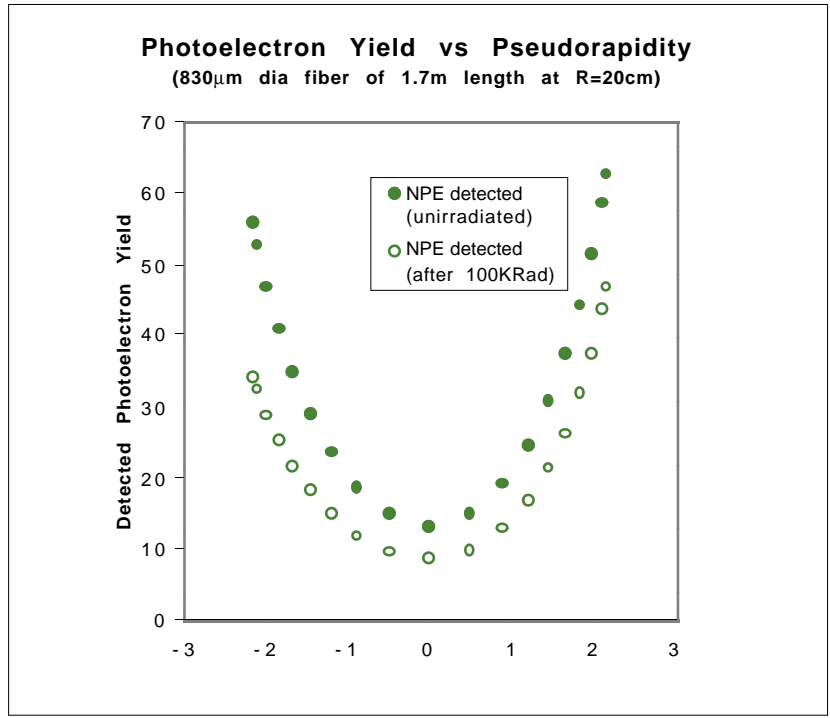,width=4.3in}}
\caption{Detected photoelectron signal for the inner layer of the CFT  as
a function of pseudorapidity, before and after 100~krad of irradiation.  
This integrated dose corresponds to 1 TeV33 year.}
\label{fig:randy2}
\end{figure}
 	
\begin{figure}[hp]
\centerline{\psfig{figure=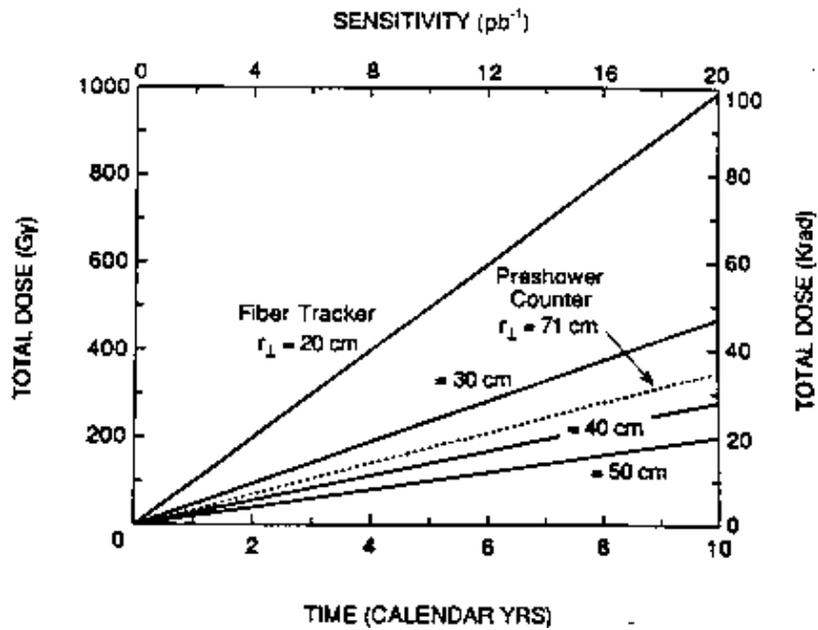,width=4.3in}}
\caption{Radiation dose due to charged particle fluence as a 
function of radial position, in Run~II years. 
(Ten Run~II years = one TeV33 year.)}
\label{fig:randy3}
\end{figure}

\begin{figure}[hp]
\centerline{\psfig{figure=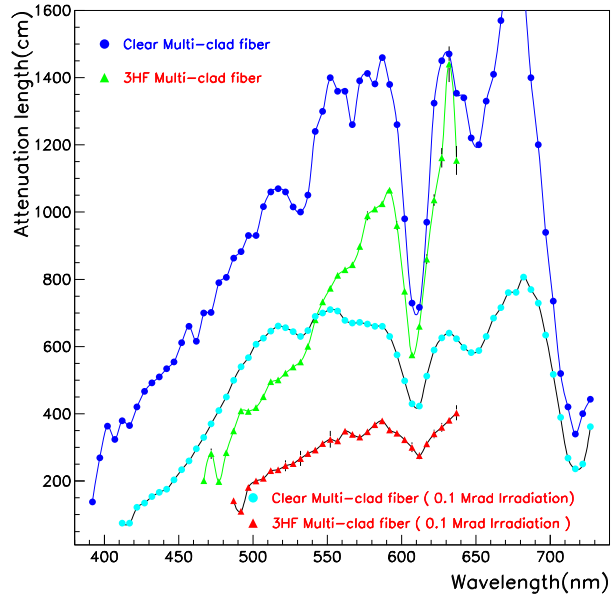,width=3.5in}}
\caption{ Optical transmission as a function of 
wavelength for 3HF scintillating fiber 
and clear waveguide fiber before and after 100 krad of irradiation.}
\label{fig:randy4}
\end{figure}

\subsection{Radiation Damage to the Scintillating Fibers}

The effect of radiation damage~\cite{dose} is approximately
proportional to $r^{-2}$. Hence, detectors placed at the inner radii are
most at risk, see Fig.~\ref{fig:randy3}.  In one TeV33 year, we expect
an integrated dose of approximately 100~krad to scintillating fibers
situated on the innermost support cylinder ($r$~=~20 cm) of the CFT.
This corresponds to a factor of two reduction in the optical
attenuation length of these fibers, see Fig.~\ref{fig:randy4}.
Additionally, $\sim$1.5 m of the clear fiber waveguide will be
similarly affected.  Together, these effects result in a 33\% drop in
the detected photoelectron yield at $\eta \sim$ 0 from fibers placed
at $r = 20$~cm, as indicated in Fig.~\ref{fig:randy2}.
 
The light loss reduces the tracking efficiency for an inner
fiber-doublet layer from a value $>$99\% to $\simeq$98\%.  While
designed to survive for 10 years of operation in Run~II, the 
inner scintillating fiber layers would need to be moved to  larger
radii for Run~III.

\begin{figure}[hp]
\centerline{\psfig{figure=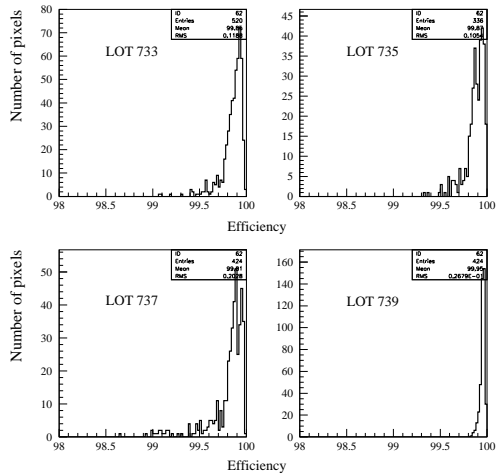,width=2.8in}}
\caption{Efficiency of about 1600 channels of VLPCs at an
equivalent single photoelectron rate of 40~MHz, typical of TeV33
operation at inner tracking radii.  Efficiencies greater than 98.5\%\
are indicated. }
\label{fig:randy5}
\end{figure}

\subsection{ Rate Dependence of the VLPCs}  
	
Potential rate dependences of the gain and quantum efficiency of the
visible light photon counters have been studied for 1600 channels of
the devices to be used in Run~II.  By operating them at a temperature
of $\sim$ 9K and at a nominal 6.5 V bias, full performance can be
maintained at photorates well above those expected for either Run~II
or Run~III.  Figure~\ref{fig:randy5} displays the efficiency of these
devices at 40~MHz.  The rate dependence leads to a reduction in
detection efficiency per fiber doublet ribbon from 99\% to 98\% up to
the maximum luminosity expected.

\subsection{ Increased Occupancy of the Fiber Layers.}

Tracking detector occupancy is proportional to $\overline{n}$, see
Fig.~\ref{fig:randy6}.  Simulation studies have shown
(Fig.~\ref{fig:randy7}) that ghost clusters in the tracking layers and
Level 1 tracking-trigger rates increase as a strong function (more
than linear) of $\overline{n}$.  This occurs because the dominant
contribution to tracking triggers at fixed $p_T$ threshold are
fakes (accidental alignments of combinations of hits from lower $p_T$
tracks) particularly in jets.  To operate within the 10~kHz trigger
bandwidth available into Level 2, it is essential to maintain a useful
Level 1 track trigger in combination with calorimeter, preshower, and
muon-system triggers.  This necessitates keeping $\overline{n}$ as
small as possible and also keeping the occupancy per fiber as small as
possible.  Luminosity leveling can control $\overline{n}$.
Minimizing the occupancy per fiber requires either modification or
repositioning of the fiber layers at larger radii.

\begin{figure}[hp]
\centerline{\psfig{figure=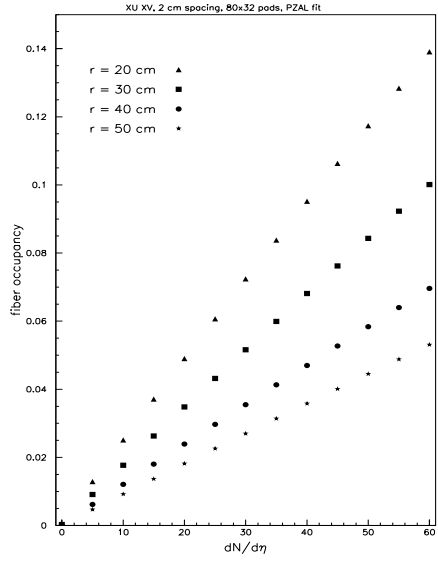,height=2.2in,width=2.5in}}
\caption{Occupancy as a function of $\overline{n}$ for various tracking
radii.  Scintillating fibers of 830~$\mu$m diameter are shown.
$\overline{n}\ge 5$ (or dN/d$\eta$ $\ge 50$) is typical for TeV33
operation at 132 ns with luminosity leveling.}
\label{fig:randy6}
\end{figure}

\clearpage

\section{ Muon Trigger Rates }

 The Level 1 muon trigger for TeV33 is planned to be the same as that
 used in Run~II.  The Run~II Level 1 muon trigger searches for muons
 locally in octants in each central and forward geographic
 region. Within each octant, one muon trigger card uses combinatoric
 logic to match tracks from the CFT track trigger with either central
 or forward muon scintillation counters.  The 20~ns gate used with the
 scintillation counters provides a powerful means of rejecting
 particles associated with showers originating in beam line elements
 or detector edges. A second muon trigger card searches for muons
 using wire hit information from the central PDTs and forward MDTs.
 Wire hits are confirmed by coincidence with the corresponding
 scintillation counters.  Because the MDTs have three or four decks in
 each layer, track ``stubs'' or centroids can be formed. The angle of
 these ``stubs'' provides additional rejection of background not
 associated with the primary collisions.  The octant trigger decisions
 for each region are collected by a muon trigger crate manager which
 forms trigger decision bits for that region.  A ``loose'' muon
 trigger is defined as the coincidence of a CFT track above a given $p_T$
 threshold with muon scintillator hits found in the 20~ns gate. A
 ``tight'' muon trigger requires additionally, centroids found using
 muon chamber hit information.  Regional trigger decisions from the
 muon trigger crate managers are sent to the muon trigger manager
 which forms a global muon trigger decision which is sent to the
 trigger framework for inclusion in the global physics trigger.

\begin{figure}[h] \centering
\mbox{}
\epsfysize=4.0in 
\epsfbox{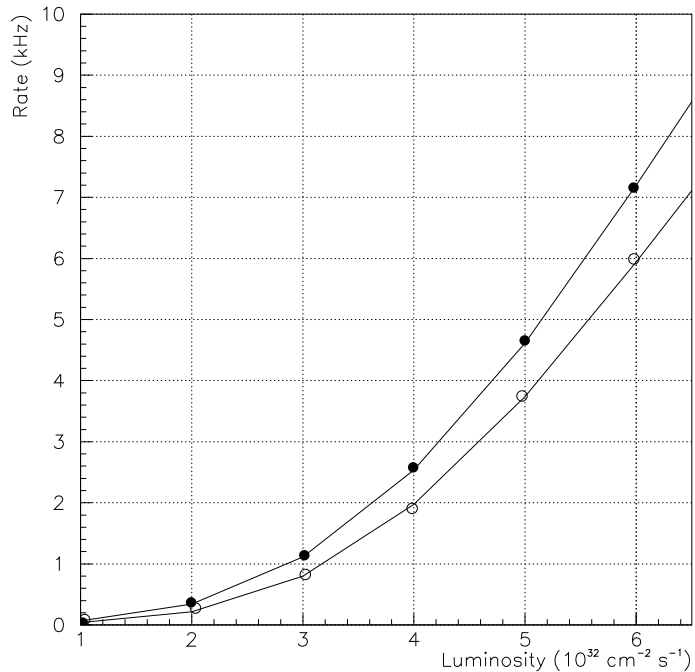}
\caption{Estimated muon trigger rates as a function of luminosity. The high
$p_T$ single muon trigger is shown as the open circles and the low
$p_T$ dimuon trigger as the filled circles. The muons are ``loose'' as
defined in the text.}
\label{muon_trigger_rates}
\end{figure}

 Muon trigger rates for TeV33 have been estimated using {\sc isajet}
 two-jet event samples generated with 1, 2, 4, and 6 interactions per
 crossing.  The events were fully simulated in the Run~II detector
 using {\sc geant}.  A simulation of the muon trigger logic was used
 to find the pass rate as a function of the number of interactions per
 crossing.  An extrapolation of the pass rate was made to estimate the
 rate for the $\approx 20\%$ of events with $\ge 7$ interactions per
 crossing.  There is an uncertainty on the estimated trigger rate of
 approximately a factor of two due to uncertainties in how well the
 {\sc isajet/geant} simulation reflects actual TeV33 conditions and to limited
 Monte Carlo statistics.

 A goal of the design of the Run~II muon system upgrade is to have two
 basic triggers which are unprescaled: a high $p_T$ single muon
 trigger and a low $p_T$ dimuon trigger.  The former is the ``$W/Z$''
 trigger which is also used for top physics and new phenomena
 searches.  The latter is a ``$b$'' physics trigger which also
 collects a large sample of $\psi$'s for calibration.  The background
 rate for the ``loose'' version of these triggers is shown in
 Fig.~\ref{muon_trigger_rates} as a function of luminosity.  To set
 the scale for the rate due to actual physics processes, the single
 muon rate from $b\bar{b}$ production alone is roughly 100 Hz at the
 TeV33 luminosity.  For Run~II, the rate for both triggers is $<500$
 Hz but the effect of multiple interactions is clearly visible in
 Fig.~\ref{muon_trigger_rates} from the nonlinear rise in rates with
 luminosity.  These early studies indicate that the ``loose'' triggers
 may be problematic for TeV33.  The additional rejection, however,
 available from the ``tight'' trigger should be sufficient to control
 the rates for TeV33.  In particular, the ``tight'' single muon
 trigger has a rate well below 100~Hz at a luminosity of
 $5\times10^{32}~{\rm cm}^{-2}{\rm s}^{-1}$ compared with 4 kHz for
 the ``loose'' single muon trigger.  The conclusion is that the muon
 trigger rates will, with minimal efficiency loss, stay within their
 allotted Level~1 bandwidth for TeV33.


\end{document}